\documentclass[aps,prfluids,onecolumn,longbibliography,superscriptaddress,amsmath,amssymb]{revtex4-2}

\usepackage{bm}

\usepackage[dvips]{graphicx}
\usepackage[dvips]{color}

\begin{document}

\title{
Variation of focusing patterns of laterally migrating particles
in a square-tube flow
due to non-Newtonian elastic force
}

\author{Naoto Yokoyama}
\affiliation{Department of Mechanical Engineering, Tokyo Denki University, Adachi 120-8551, Japan}
\author{Hiroshi Yamashita}
\affiliation{Department of Pure and Applied Physics, Kansai University, Suita 564-8680, Japan}
\affiliation{Department of Mechanical Science and Bioengineering, Osaka University, Toyonaka 560-8531, Japan}
\author{Kento Higashi}
\affiliation{Department of Pure and Applied Physics, Kansai University, Suita 564-8680, Japan}
\author{Yuta Miki}
\affiliation{Department of Pure and Applied Physics, Kansai University, Suita 564-8680, Japan}
\author{Tomoaki Itano}
\affiliation{Department of Pure and Applied Physics, Kansai University, Suita 564-8680, Japan}
\author{Masako Sugihara-Seki}
\affiliation{Department of Pure and Applied Physics, Kansai University, Suita 564-8680, Japan}
\affiliation{Department of Mechanical Science and Bioengineering, Osaka University, Toyonaka 560-8531, Japan}

\date{\today}

\begin{abstract}
The elasto-inertial effects on particle focusing in a square-tube flow were investigated experimentally and numerically.
Microscale experiments using spherical particles in dilute polymer solutions demonstrated that
the particles are focused on the midline and/or the diagonal in a downstream cross-section,
depending on the polymer concentration.
Numerical computations based on the FENE-P model for the viscoelastic flow
reproduced these focusing patterns.
It was revealed that
the transitions among the patterns are accounted for by
the elastic forces due to the first normal stress difference and the polymer elongation,
which are the essentials of the viscoelastic fluid.
\end{abstract}

\maketitle

Lift due to inertia causes particles suspended in a tube flow
to travel the streamwise direction
and to migrate in the lateral direction perpendicular to the mainstream.
This phenomenon is called the Segr\'e--Silberberg (SS) effect~\cite{segre_silberberg_1962}.
Inertial particles in circular-tube flows
appear only at the so-called SS annulus in the downstream cross-section
at small Reynolds numbers $\mathrm{Re}$.
The radius of the SS annulus is roughly $0.6$ times as large as the tube radius,
and is determined by a balance between the lift due to the shear gradient and the wall effect~\cite{matasmorrisguazzelli,*ho_leal_1976}.
On the other hand,
in square-tube flows, which lack the axial symmetry,
inertial particles appear at isolated focusing positions instead of the SS annulus located
on the midline, diagonal, and others~\cite{doi:10.1063/1.4902952,PhysRevFluids.2.044201,doi:10.1063/1.2176587,*miura_itano_sugihara-seki_2014,*DiCarlo18892,*C0LC00212G}.
Various focusing patterns emerge,
depending on $\mathrm{Re}$ and the blockage ratio $\kappa = d/W$,
where $d$ and $W$ respectively represent the particle diameter and the tube width~\cite{PhysRevFluids.2.044201,PhysRevFluids.4.124307,PhysRevLett.102.094503,C4LC00145A}.
For deformable particles such as living cells, their deformability also affects their focusing positions~\cite{C0LC00595A,*doi:10.1063/1.3664402}.
If the focusing positions can be controlled by the size, shape, and rheological properties such as the deformability of the suspended particles,
the SS effect can be used for particle separation and sorting.
To this end, extensive studies have been performed to apply the SS effect to a suspension flow of living cells and particles in microfluidics.
The development of continuous and easy separation as well as the purification of living cells without damage is strongly required
especially in life science and medical care including clinical practice~\cite{Davis14779}.

The medium where living cells are suspended
usually contains polymers such as proteins and has viscoelasticity.
In a viscoelastic tube flow,
suspended particles experience the lift due to elasticity,
which points inwards except near the corner in the cross-section~\cite{LU2017182}.
Even in an almost Newtonian fluid medium with a short relaxation time,
the large shear strain can cause a notable non-Newtonian behavior~\cite{Kimeaav4819}.
Moreover,
the controllability of the focusing characteristics
by the rheological properties of the medium
has also been reported~\cite{PhysRevLett.98.234501}.

Most previous studies on the particle migration in viscoelastic flows concern the particle focusing on the tube center and the tube corner~\cite{li_mckinley_ardekani_2015,doi:10.1063/1.4882265,C0LC00102C,*C2LC21304D,*C2LC21154H,*DelGiudice2015}.
However, as flow rates increase to realize a high throughput,
inertial effects on the particle migration are enhanced as $\mathrm{Re}$ increases.
Since the inertial lift due to the shear gradient directs outwards
as opposed to inward lift due to elasticity,
the particle focusing position should vary owing to the interplay between inertia and elasticity.
In fact, a recent numerical study based on the Oldroyd-B model predicted that
suspended spherical particles in square-tube flows focus
on the midline and diagonal as well as at the tube center and tube corners,
depending on the Weissenberg number $\mathrm{Wi}$~\cite{yu_wang_lin_hu_2019}.
On the other hand,
few experimental observations have reported the focusing of particles at the intermediate positions between the tube center and the tube wall.
Thus, the combined effects of the inertia and elasticity of the media on the focusing remain unclear yet.

Here,
we experimentally investigate the migration of spherical particles suspended in square-tube flows of dilute polymer (polyvinylpyrrolidone, PVP) solutions
to demonstrate the presence of intermediate focusing positions of particles.
To explain the focusing patterns in the viscoelastic flow,
we also performed numerical computations for a neutrally buoyant spherical particle in viscoelastic fluids flowing through a square tube
based on the finitely extensible nonlinear elastic-Peterlin (FENE-P) model~\cite{https://doi.org/10.1002/pol.1966.110040411,*BIRD1980213}.
The lateral force due to the elasticity acting on the particle is quantitatively evaluated,
and the transition of the focusing pattern is rationalized in terms of the first normal stress difference and the polymer elongation near the tube wall.

\begin{figure}
 \centering
\includegraphics[width=.95\linewidth]{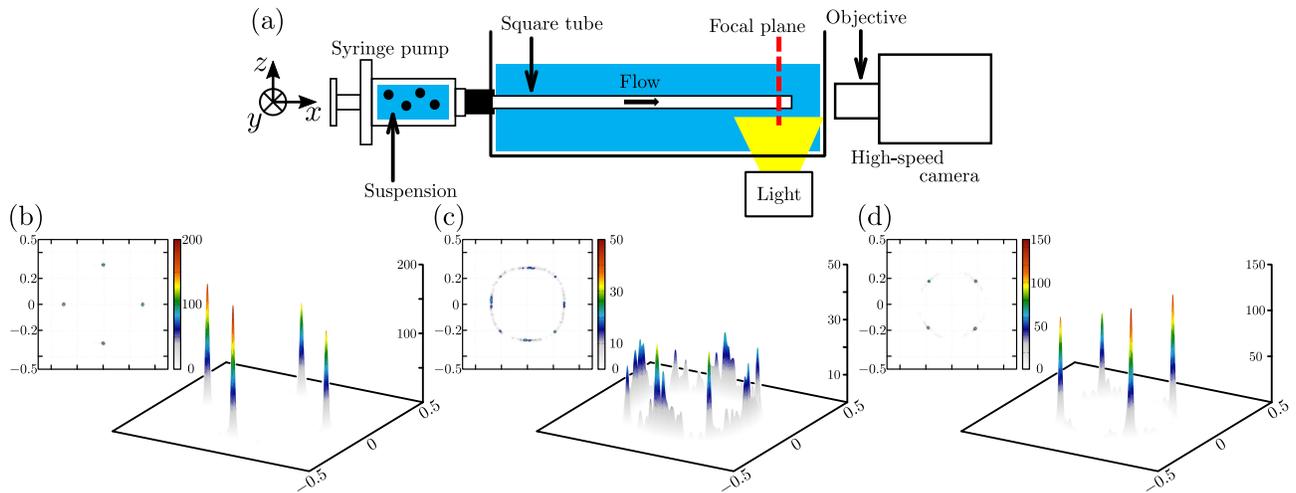}
 \caption{
 Focusing positions in the PVP-solution experiments.
(a) Experimental setup (schematic).
(b)--(d)
Existence probability of particles on a cross-section near the outlet of the tube
with the PVP concentrations
(b) $\phi = 1$~wt\%,
(c) $\phi = 1.9$~wt\%,
and
(d) $\phi = 2.5$~wt\%.
($\mathrm{Re}=50$ and $\kappa = 0.15$)
}
\label{fig:expstreamwise}
\end{figure}
In the laboratory experiment,
spherical polystyrene particles dilutely suspended in aqueous solutions of PVP, biocompatible synthetic polymers, flowing through a square tube
were observed to detect their positions in a cross-section.
Figure~\ref{fig:expstreamwise}(a) shows the experimental system
(see Ref.~\cite{PhysRevFluids.2.044201} for details).
The solutions contain PVP (Mw = 360kDa, Tokyo Chemical Industry) at the concentrations of $\phi = 1$, $1.9$, and $2.5$~wt\%,
whose viscosities are $\eta = 3.2$, $7.4$, and $10$~mPa\,s at $22$~$^{\circ}$C, respectively.
These viscosity values were measured using a rotational rheometer (Haake Mars III) and are nearly independent of the shear rate.
Their elasticities are too small to evaluate the relaxation time from the dynamic moduli,
and hence $\mathrm{Wi}$ in these experiments could not be evaluated.
Since the PVP solution is dilute,
the non-Newtonian properties are quite weak.
Spherical polystyrene particles (diameter $d = 60$~$\mu$m)
were suspended in the PVP solution with volume fractions of order $10^{-2}$~\%.
A syringe pump (Nexus 6000, ISIS) was used to infuse the suspension
through a straight glass duct (VitroCom) with a square cross-section of the width $W = 400$~$\mu$m and a length $L = 600$~mm.
The syringe and the duct were connected via a circular tube with inner diameter of $1.14$~mm, and the contraction ratio is approximately $1/3$.
Moreover, the Poiseuille flow is built roughly at $x/W \approx \mathrm{Re}/30 \approx 1.7$
and the entrance effects are negligible.
More than $500$ particles were imaged in the cross-section near the outlet of the tube
with a high-speed camera (AX100, Photron) equipped with 20$\times$ objective,
and the particle positions were determined by image analysis (Image J~\cite{Schneider2012}, NIH).
The Reynolds number was set to be $\mathrm{Re} = UW/\nu = 50$,
which results in the particle Reynolds number $\mathrm{Re}_{\mathrm{p}} = \mathrm{Re}\, \kappa^2 = 1.125$.
The blockage ratio is $\kappa = 0.15$.
Here, $U$ and $\nu$ are the mean velocity and the kinematic viscosity,  respectively.
The $x$-axis corresponds to the centerline of the tube,
and  the $y$- and $z$-axes,
which are normal to the tube walls,
span the cross-section,
In addition to the Cartesian coordinate $(y,z)$,
the polar coordinate $(r,\theta)$ is also used in the cross-section.

Figures~\ref{fig:expstreamwise}(b)--(d) show
the existence probability of particles on a cross section near the outlet of the tube
for various PVP concentrations.
In the most dilute PVP solution ($\phi = 1$~wt\%, Fig.~\ref{fig:expstreamwise}(b)),
the particles are focused on the midlines of the cross-section
similar to Newtonian flows~\cite{PhysRevFluids.2.044201,PhysRevLett.102.094503}.
These focusing positions are referred to as the midline equilibrium position (MEP).
As the PVP concentration increases,
new focusing positions,
which are referred to as the diagonal equilibrium positions (DEP),
emerge on the diagonal for $\phi = 1.9$~wt\% (Fig.~\ref{fig:expstreamwise}(c)).
For $\phi = 1.9$~wt\%, the MEP and the DEP are bistable,
and the particles are focused near these eight equilibrium positions.
Some particles do not reach the focusing positions completely yet,
and the ratio $L/W=1500$ might not be large enough for $\phi = 1.9$~wt\%.
\footnote{
The numerical computation evaluates the streamwise distance for which the particle is swept downstream as $x/W \sim O(10^3)$
during the migration along the circular ring on the cross-section
because the mean lateral force along the ring is roughly evaluated as $4\times 10^{-6}$ there,
and the mean migration velocity is determined by the quasi-steady balance
between the lateral lift and the Stokes drag.
}
However,
because the existence probabilities of the particles at the focusing positions
are much larger than those out of the focusing positions,
the focusing positions from these experiments are clearly identified.
Further increasing the PVP concentration eliminates the MEP,
and the particles are focused only near the DEP for $\phi = 2.5$~wt\% (Fig.~\ref{fig:expstreamwise}(d)).
Figures~\ref{fig:expstreamwise}(b)--(d) demonstrate that
the three focusing patterns and the transitions among them emerge for $\mathrm{Re} = 50$ and $\kappa = 0.15$
as the PVP concentration increases.
It is noteworthy that particle focusing on the tube center is also observed in higher-concentration PVP solutions ($\phi = 8$~wt\%) at low $\mathrm{Re}$.
This is consistent with the previous study~\cite{doi:10.1063/1.4882265},
but the result is not shown here.

The Weissenberg number $\mathrm{Wi}$ is defined as the ratio of the elastic force to the viscous force
and is considered to increase as the PVP concentration increases.
Numerical computations by using the FENE-P model for the viscoelastic flow
were performed to find the lift distribution and the focusing positions,
which are sinks of the vector field of the lift,
depending on $\mathrm{Wi}$ in comparison with the above experiment.

The governing equation nondimensionalized by the mean-flow velocity and the tube width
is written as
\begin{align}
&
 \frac{\partial \bm{u}}{\partial t} + (\bm{u} \cdot \nabla) \bm{u}
 = - \nabla p + \frac{\beta}{\mathrm{Re}} \nabla^2 \bm{u} + \nabla \cdot \mathsf{\tau}_{\mathrm{p}}
 + \bm{f}_{\mathrm{IB}}
 + \bm{f}_{\mathrm{PG}}
 ,
 \label{eq:NS}
   \\
&
   \nabla \cdot \bm{u} = 0
,
  \end{align}
where $\bm{u}$, $p$, and $\mathsf{\tau}_{\mathrm{p}}$ are respectively
the velocity of the flow, the pressure, and the viscoelastic stress tensor derived from the polymers.
Moreover,
$1/\mathrm{Re}$ is the nondimensionalized total shear viscosity.
The ratio of the shear viscosity due to the solvent Newtonian fluid to the total viscosity
is expressed by $\beta$,
and $\beta=1/2$ in this Letter.
Note that
the non-dimensional lateral velocity is smaller than $10^{-6}$ in the absence of the particle at $\mathrm{Re}=50$ in the $\mathrm{Wi}$ range examined in this Letter,
and the secondary flow is negligible,
since the non-Newtonian properties are weak.
The interaction between the inertial particles and the flow is modeled by the volume force $\bm{f}_{\mathrm{IB}}$
according to the immersed boundary method~\cite{kajishima2001}.
The flow is driven by the mean pressure gradient $\bm{f}_{\mathrm{PG}}$,
so that the volume flux is constant.
To compare the focusing positions in the experiments and the numerical computations,
$\mathrm{Re}=50$ and $\kappa=0.15$ are employed.

The viscoelastic fluid of the dilute polymer solutions is modeled by the FENE-P model,
where the viscoelastic stress tensor $\mathsf{\tau}_{\mathrm{p}}$
is given by the positive-definite symmetric conformation tensor $\mathsf{C}$:
\begin{align}
 \mathsf{\tau}_{\mathrm{p}} = \frac{1-\beta}{\mathrm{Re}\mathrm{Wi}} \left(\frac{\mathsf{C}}{1 - \mathrm{tr}(\mathsf{C})/l^2} - \mathsf{I}\right)
.
\end{align}
Here, the maximal length of the polymers $l$ is set to $10$.
The square-root symmetric tensor $\mathsf{B}$ instead of $\mathsf{C}=\mathsf{B}^2$
is employed for numerical stability~\cite{BALCI2011546},
and the governing equation of $\mathsf{B}$ is written as
\begin{align}
 \frac{\partial \mathsf{B}}{\partial t}
 =& - \bm{u} \nabla \mathsf{B}
 + \mathsf{B} \nabla \bm{u}
 + \mathsf{A} \mathsf{B}
 + \frac{1}{2\mathrm{Wi}} \left(\mathsf{B}^{-1} - \frac{\mathsf{B}}{1 - \mathrm{tr}(\mathsf{B}^2)/l^2}\right)
,
 \label{eq:FENEPsq}
\end{align}
where
$\mathsf{A}$ is the antisymmetric tensor that satisfies
$\mathsf{A}\mathsf{B} + \mathsf{B}\mathsf{A} = (\nabla \bm{u})^{\mathrm{T}} \mathsf{B} - \mathsf{B} \nabla \bm{u}$.

These governing equations are numerically integrated by a fractional-step method.
The Adams--Bashforth method is used for the convection term and the non-Newtonian viscous term
in the Navier--Stokes equation~(\ref{eq:NS}),
and the Newtonian viscous term is implicitly added.
The mean pressure gradient is determined by the parameter adaptive control
so that the mean velocity is unity.
The Poisson equation for the pressure is solved by the biconjugate gradient stabilized (BiCGStab) method or the successive over relaxation (SOR) method.
The Adams--Bashforth method is used also for the governing equation of $\mathsf{B}$~(\ref{eq:FENEPsq}) and that of the particle motion~(\ref{eq:particlemotion}) below.
The periodic boundary condition is used in the streamwise direction
with a period length $L_x = 2$.
That is, the period length is twice as long as the tube width,
and more than $13$ times longer than the particle's diameter.
The relative difference of the lateral force acting on the particle placed at $(y_{\mathrm{C}},z_{\mathrm{C}}) \approx (0.174,0)$
at $\mathrm{Wi}=0.16$
between $L_x=1$ and $L_x=4$ is approximately $3$\%,
and that between $L_x=2$ and $L_x=4$ is approximately $1$\%.
These small relative differences validate the period length $L_x = 2$ as well as the periodic boundary condition.

The volume force by using the immersed boundary method is given as
\begin{align}
\bm{f}_{\mathrm{IB}} = \frac{\alpha}{\Delta t} (\bm{U} - \bm{u}^{\ast})
,
\end{align}
where $\bm{U}$ and $\bm{u}^{\ast}$ are respectively
the particle's rigid-body velocity at the center of each computational cell near the particle and the intermediate velocity of the flow~\cite{kajishima2001}.
Here, $\alpha$ represents the volume fraction of the inertial particles in the computational cell.
The force and the torque acting on the particle are respectively
the reaction force and torque of the volume force,
and obtained as
\begin{align}
 \bm{F}_{\mathrm{3D}} = \int (-\bm{f}_{\mathrm{IB}}) dV,
 \qquad
 \bm{N} = \int (\bm{x} - \bm{x}_{\mathrm{C}}) \times (-\bm{f}_{\mathrm{IB}}) dV
.
\end{align}
The projection of the three-dimensional force $\bm{F}_{\mathrm{3D}}$ onto the cross-section,
that is, the lift acting on the particle $\bm{F}$
is investigated in this Letter.
The particles are fixed in the lateral direction
while they travel in the streamwise direction and rotate freely.
Thus,
the equation of motion
for the streamwise velocity $U_{\mathrm{C}x}$ of the center of mass of the particle $\bm{x}_{\mathrm{C}}$
and the angular velocity $\bm{\Omega}_{\mathrm{C}}$
is written as
\begin{align}
 m \frac{d U_{\mathrm{C}x}}{dt} = F_{\mathrm{3D}x},
\qquad
 I \frac{d \bm{\Omega}_{\mathrm{C}}}{dt} = \bm{N}
.
\label{eq:particlemotion}
\end{align}
The mass and the inertial moment of the particles
are expressed by $m=\pi \kappa^3/6$ and $I=m\kappa^2/10$, respectively.

\begin{figure}
 \centering
\includegraphics[width=.8\linewidth]{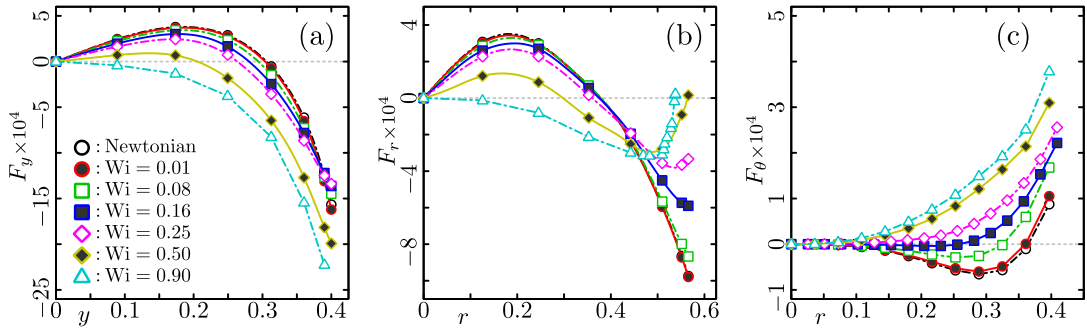}
 \caption{
 Radial component of the lift exerted on the particle
 located (a) on the midline and (b) on the diagonal.
 Azimuthal component of the lift exerted on the particle
 located (c) on the line $z=y \tan(\pi/8)$.
}
 \label{fig:forceonlines}
\end{figure}

Figures~\ref{fig:forceonlines}(a)--(b) plot
the radial component of lift $F_r$
acting on the particle placed on the midline ($y$-axis) and the diagonal.
Note that $F_r = F_y$ on the $y$-axis.
The lift in Newtonian flows is directed outward, i.e., $F_r>0$ near the tube center
while $F_r<0$ near the tube wall~\cite{PhysRevLett.102.094503,nakagawa_yabu_otomo_kase_makino_itano_sugihara-seki_2015}.
As $\mathrm{Wi}$ increases, $F_r$ decreases
except on the diagonal near the tube wall, i.e., tube corners
($r\gtrapprox 0.5$ in Fig.~\ref{fig:forceonlines}(b)),
and
the zeros of $F_r$ slightly move closer to the origin
in the range $0\leq\mathrm{Wi}\leq 0.25$.
Note that these zeros correspond to the MEP (Fig.~\ref{fig:forceonlines}(a)) and the DEP (Fig.~\ref{fig:forceonlines}(b))
because $F_{\theta}=0$ on these lines.
These zeros of $F_r$ disappear because $F_r<0$ except near the corners for the large $\mathrm{Wi}$ ($= 0.9$).
This result implies that the inward lift acts on the particle at large $\mathrm{Wi}$
similar to previous studies~\cite{doi:10.1063/1.4882265,yu_wang_lin_hu_2019}.

For a large $\mathrm{Wi}$ ($= 0.9$),
$F_r < 0$ at $r\lessapprox 0.52$ while $F_r > 0$ at $r\gtrapprox 0.52$
on the diagonal (Fig.~\ref{fig:forceonlines}(b)).
Thus, the tube center becomes a stable equilibrium position,
and another focusing position appears near the corner outside of the saddle position where the particle almost touches the two walls.
Therefore, the corner equilibrium positions as well as the center are predicted to be focusing positions for $\mathrm{Wi}\gtrapprox 0.9$.

Similarly,
Fig.~\ref{fig:forceonlines}(c) plots
the azimuthal component of lift $F_{\theta}$
acting on the particle placed on a line of $z=y\tan(\pi/8)$.
The azimuthal lift
$F_{\theta}$ increases monotonically
and the zeros of $F_{\theta}$ approach the origin
with increasing $\mathrm{Wi}$.
At the large $\mathrm{Wi}$ ($=0.9$)
these zeros vanish with $F_{\theta}>0$
on a line of $z=y\tan(\pi/8)$.
This result indicates that the lift acts towards the diagonal at large $\mathrm{Wi}$,
because the sign of $F_{\theta}$ in $\pi/4<\theta<\pi/2$ is opposite to that in $0<\theta<\pi/4$ from the symmetry.
As shown in Fig.~\ref{fig:FEfnsd} below,
the lift toward the diagonal is one of the most important features of the elastic lift.

\begin{figure}
 \centering
\includegraphics[width=.8\linewidth]{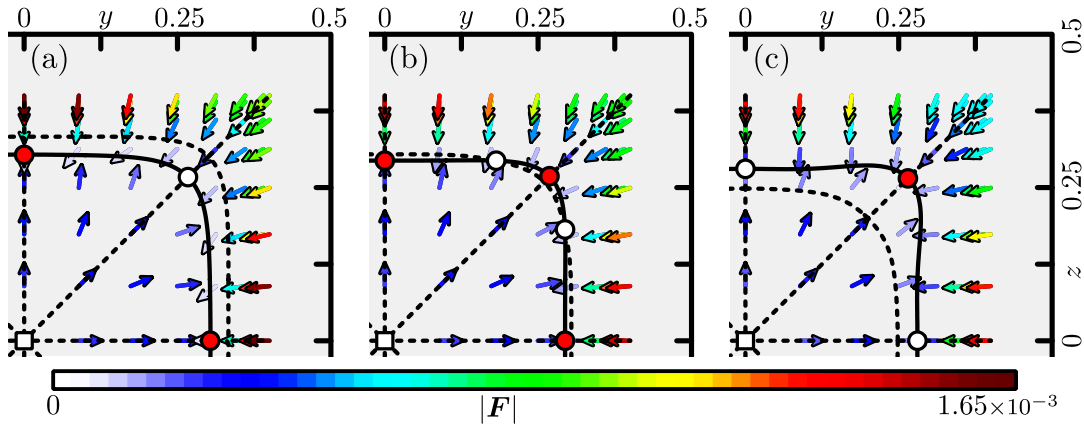}
 \caption{
Lift and equilibrium positions.
Stable, saddle, and unstable equilibrium positions are respectively represented by
\textcolor[named]{Red}{{\Large $\bullet$}}, $\bigcirc$, and $\Box$.
 The solid line and the dashed lines
 represent the contour of $F_r=0$,
 and those of $F_{\theta}=0$, respectively.
Each arrow represents the lift direction with its magnitude shown by color.
 (a) $\mathrm{Wi}=0.01$, (b) $\mathrm{Wi}=0.08$,
 and (c) $\mathrm{Wi}=0.16$.
}
\label{fig:lateralforce}
\end{figure}

Figure~\ref{fig:lateralforce} shows
the lift in the first quadrant of the cross-section
for $\mathrm{Wi}=0.01$, $0.08$,
and $0.16$.
Stable, saddle, and unstable equilibrium positions are also displayed,
and the focusing positions observed experimentally correspond to the stable equilibrium positions.

For the smallest $\mathrm{Wi}$ ($=0.01$),
the lift in Fig.~\ref{fig:lateralforce}(a) is quite similar to that obtained
in Newtonian flows~\cite{PhysRevFluids.4.124307,PhysRevLett.102.094503,nakagawa_yabu_otomo_kase_makino_itano_sugihara-seki_2015}.
The contour of $F_r =0$ ($r$-nullcline) forms a closed loop almost parallel to the walls between the tube center and the tube walls.
On the other hand,
the contours of $F_{\theta}=0$ consist of the midlines ($y$- and $z$-axes), the diagonal,
and a closed loop just outside of the $r$-nullcline.
Although all the contours of $F_{\theta}=0$ are nullclines,
here we refer only to the closed loop as $\theta$-nullcline for convenience.
Note that $F_r>0$ inside the $r$-nullcline whereas $F_r<0$ outside it.
Similarly, $F_{\theta}<0$ inside the $\theta$-nullcline
whereas $F_{\theta}>0$ outside it
in $0<\theta <\pi/4$,
and vice versa in $\pi/4<\theta <\pi/2$.

From the symmetry, the tube center is always an equilibrium position,
where the lift vanishes.
The lift direction around the tube center indicates that
the tube center is an unstable equilibrium position for $\mathrm{Wi}=0.01$.
The intersections between the zero contours of $F_r$ and $F_{\theta}$ represent the equilibrium position;
the intersection between the $r$-nullcline and the midline ($y$- and $z$-axes)
correspond to the MEP,
and that between the $r$-nullcline and the diagonal does to the DEP.
At $\mathrm{Wi}=0.01$, the MEP is stable,
but the DEP is a saddle, judged from the lift direction around them.
This predicts that the particle focuses on the MEP at small $\mathrm{Wi}$,
which agrees with the experiment of $\phi=1$~wt\% in Fig.~\ref{fig:expstreamwise}(b).

At $\mathrm{Wi}=0.08$ in Fig.~\ref{fig:lateralforce}(b),
the $\theta$-nullcline approaches the tube center,
whereas the $r$-nullcline does not change its position so much.
Thus, these two nullclines cross,
and the intersection represents another equilibrium position.
The lift indicates that
this new equilibrium position is a saddle, and both MEP and DEP are stable.
This result is consistent with the experiment of $\phi=1.9$~wt\% in Fig.~\ref{fig:expstreamwise}(c).

A further increase in $\mathrm{Wi}$ moves the $\theta$-nullcline closer to the tube center.
At $\mathrm{Wi}=0.16$,
the $\theta$-nullcline exists inside the $r$-nullcline (Fig.~\ref{fig:lateralforce}(c)).
In this case,
the MEP is unstable while the DEP is stable
in accordance with the experiment of $\phi=2.5$~wt\% in Fig.~\ref{fig:expstreamwise}(d).

The numerical results for $\mathrm{Wi}\leq 0.16$ in Figs.~\ref{fig:lateralforce}(a)--(c) reproduce
the focusing patterns obtained experimentally for $\phi\leq 2.5$~wt\% (Figs.~\ref{fig:expstreamwise}(b)--(d)).
In the numerical computations,
these patterns transition
as $\mathrm{Wi}$ increases,
which is proportional to the relaxation time $\lambda$ of the medium.
Although $\lambda$ of the polymer solutions used in the experiments is currently difficult to measure,
$\lambda$ is reasonably assumed to increase with increasing $\phi$.
Thus, the present numerical computations explain both the focusing patterns
and the transition among them shown in Fig.~\ref{fig:expstreamwise}.

Figure~\ref{fig:lateralforce} demonstrates that
the transition of the particle focusing patterns is due to
the immobility of the $r$-nullclines and the shrinkage of the $\theta$-nullclines with increasing $\mathrm{Wi}$ ($\leq 0.16$).
The focusing pattern
is also consistent with that in an Oldroyd-B fluid flow~\cite{yu_wang_lin_hu_2019}.
\begin{figure}
 \centering
\includegraphics[width=.7\linewidth]{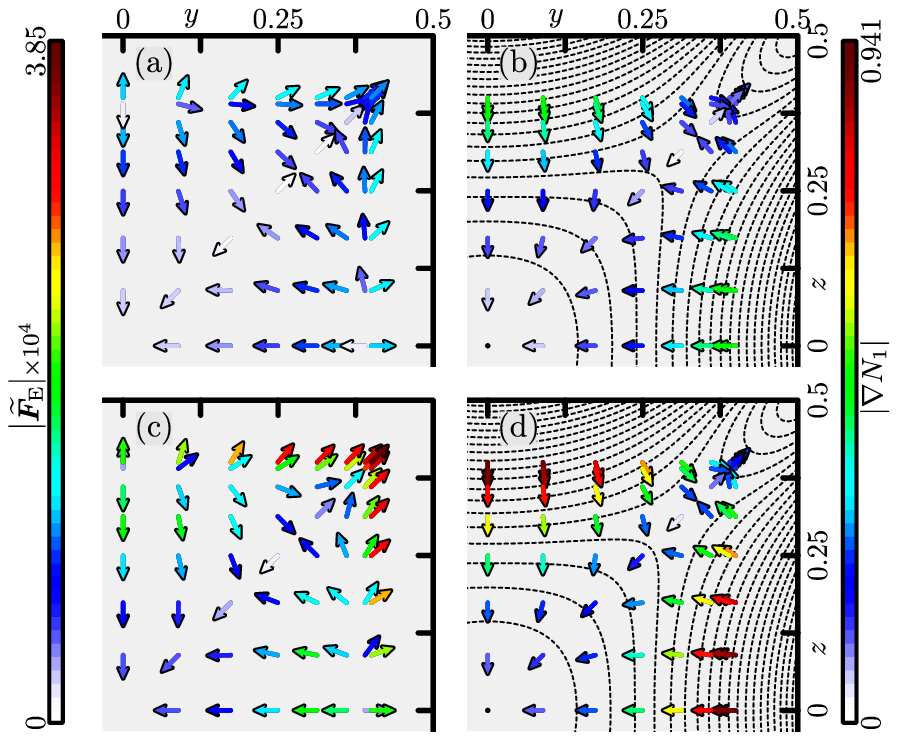}
 \caption{
 (a) and (c):
 Alternative elastic lift $\widetilde{\bm{F}}_{\mathrm{E}}=\bm{F}_{\mathrm{Wi}}-\bm{F}_0$.
(b) and (d):
 Contours of $N_1$
for every $6.85\times10^{-3}$ (b)
and $1.31\times10^{-2}$ (d).
 The arrows represent $-\nabla N_1$.
 (a)--(b) $\mathrm{Wi}=0.08$, and (c)--(d) $\mathrm{Wi}=0.16$.
 }
 \label{fig:FEfnsd}
\end{figure}
These movements of the nullclines can be understood from an increase in lift due to elasticity $\bm{F}_{\mathrm{E}}$,
which would be directed inward and toward the diagonal.
However, direct calculation of $\bm{F}_{\mathrm{E}}$ is difficult.
Instead,
$\bm{F}_{\mathrm{E}}$ is approximated by $\widetilde{\bm{F}}_{\mathrm{E}}=\bm{F}_{\mathrm{Wi}}-\bm{F}_0$,
where $\bm{F}_{\mathrm{Wi}}$ and $\bm{F}_0$
are the lifts acting on the particle placed at the same position
in a finite-$\mathrm{Wi}$ flow and in a Newtonian flow ($\mathrm{Wi}=0$), respectively.
We also computed the first normal stress difference $N_1$
using the flow in the absence of the particle,
whose velocity profile is close to the Poiseuille flow.
The variation of $N_1$ over the particle's diameter
produces the lift $\bm{F}_{\mathrm{E} N_1} = C_{N_1} \kappa^3 \nabla N_1$
in the present nondimensionalization,
where the coefficient $C_{N_1}$ is negative~\cite{PhysRevLett.98.234501,doi:10.1063/1.4882265,LU2017182,li_mckinley_ardekani_2015}.

In fact,
$\widetilde{\bm{F}}_{\mathrm{E}}$ and $-\nabla N_1$ for $\mathrm{Wi}=0.08$ and $0.16$ are qualitatively similar,
and are mainly directed inward and toward the diagonal
in the region far from the tube wall
as illustrated in Fig.~\ref{fig:FEfnsd}.
Only near the tube wall, where the wall effect due to the elasticity is significant,
$\widetilde{\bm{F}}_{\mathrm{E}}$ is directed outward and toward the diagonal.
Thus,
the wall-induced elastic lift $\bm{F}_{\mathrm{EW}}$ is considered to be directed outward.
Near the $r$-nullcline, i.e., the contour $F_r=0$ in Fig.~\ref{fig:lateralforce},
$\widetilde{\bm{F}}_{\mathrm{E}}$ is directed toward the azimuthal direction,
and hence the radial component of $\widetilde{\bm{F}}_{\mathrm{E}}$ nearly vanishes
as seen in Figs.~\ref{fig:FEfnsd}(a) and (c).
This suggests that the radial components of $\bm{F}_{\mathrm{E} N_1}$ and $\bm{F}_{\mathrm{EW}}$ balance
near the $r$-nullcline at small $\mathrm{Wi}$.
Since the magnitude of $\bm{F}_{\mathrm{E} N_1}$ is linearly proportional to $\mathrm{Wi}$
and that of $\bm{F}_{\mathrm{EW}}$ increases with $\mathrm{Wi}$,
the radial component of $\widetilde{\bm{F}}_{\mathrm{E}}$ ($\approx\bm{F}_{\mathrm{E} N_1}+\bm{F}_{\mathrm{EW}}$) remains approximately $0$
near the $r$-nullcline
independently of $\mathrm{Wi}$.
In contrast,
the azimuthal component of $\widetilde{\bm{F}}_{\mathrm{E}}$ is directed toward the diagonal,
and its magnitude increase with $\mathrm{Wi}$ there.
Consequently,
the $\theta$-nullcline shrinks faster
while the $r$-nullcline does not change its position,
as $\mathrm{Wi}$ increases.

\begin{figure}
 \centering
 \includegraphics[width=.95\linewidth]{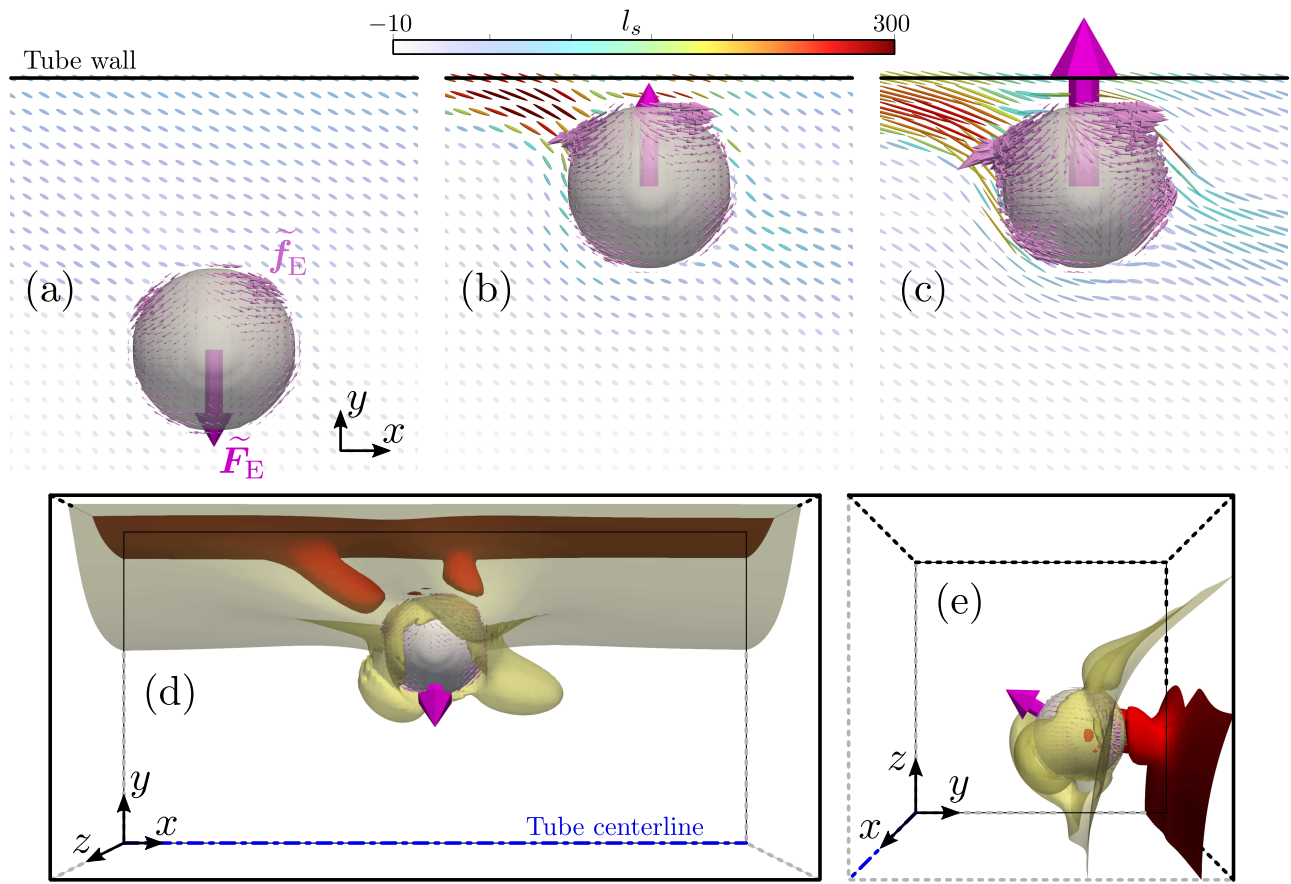}
 \caption{
 (a)--(c)
 Conformation tensor on the $z$ plane and the elastic force acting on the particle
placed at (a) $(y_{\mathrm{C}},z_{\mathrm{C}}) \approx (0.25,0)$,
and (b)--(c) $(y_{\mathrm{C}},z_{\mathrm{C}}) \approx (0.4,0)$.
 The ellipsoids colored by $l_{\mathrm{s}}$ represent the eigenvectors and eigenvalues of $\mathsf{C}$.
(d)--(e)
 Isosurfaces of $l_{\mathrm{s}}= 15$ (yellow) and $55$ (red)
 and the elastic force
when the particle center is approximately at $(y_{\mathrm{C}},z_{\mathrm{C}}) \approx (0.31,0.17)$.
 $\mathrm{Wi}=0.08$ except (c) $\mathrm{Wi}=0.16$.
}
\label{fig:conformationtensor}
\end{figure}

Next,
we consider $\widetilde{\bm{F}}_{\mathrm{E}}$ in more detail based on the polymer stretch.
While the mean-field behavior of the conformation tensor $\mathsf{C}$
determines $\bm{F}_{\mathrm{E} N_1}$,
the interaction between $\mathsf{C}$ near the wall and the particle's motion
provides $\bm{F}_{\mathrm{EW}}$.
Figures~\ref{fig:conformationtensor}(a)--(c) show
the distribution of the elastic force acting on the particle $\widetilde{\bm{f}}_{\mathrm{E}}$,
which is the difference of the interaction forces $\bm{f}_{\mathrm{IB}}$
between in a finite-$\mathrm{Wi}$ flow and in a Newtonian flow,
and $\widetilde{\bm{F}}_{\mathrm{E}}$.
 The major and minor axes of the ellipsoids represent the eigenvectors and eigenvalues of $\mathsf{C}$,
and the colors of the ellipsoids express the normalized polymer stretch defined as
$l_{\mathrm{s}} = (\mathrm{tr}\mathsf{C} -3) / (2\mathrm{Wi}^2)$~\cite{C9SM01736D}.

Since the velocity profile is close to the Poiseuille flow,
the large velocity gradient near the tube wall elongates the polymers.
The difference of $l_{\mathrm{s}}$ between near the wall and near the tube center
provides $-\nabla N_1$.
This is consistent with the observation of $\widetilde{\bm{f}}_{\mathrm{E}}$ and $\widetilde{\bm{F}}_{\mathrm{E}}$
for the particle relatively far from the wall
($(y_{\mathrm{C}}, z_{\mathrm{C}}) \approx (0.25,0)$, Fig.~\ref{fig:conformationtensor}(a));
the inward force on the outside surface $y>y_{\mathrm{C}}$ is larger than
the outward force on the inside surface $y<y_{\mathrm{C}}$,
and the resultant $\widetilde{\bm{F}}_{\mathrm{E}}$ is directed inward.
The small polymer deformation due to the interactions between the particle and the flow
results in $\widetilde{\bm{F}}_{\mathrm{E}}\approx\bm{F}_{\mathrm{E}N_1}$.

When the particle is located near the wall
($(y_{\mathrm{C}}, z_{\mathrm{C}}) \approx (0.4,0)$, Fig.~\ref{fig:conformationtensor}(b)),
the strong elongation near the wall extends to the particle from the upstream,
which would generate $\bm{F}_{\mathrm{EW}}$.
The polymers elongated in the direction normal to the particle
touch the particle on its upstream and outer side.
The strongly elongated polymers generate tension,
pulling the particle to the upstream and outward to the wall.
Therefore, the inward force of $\widetilde{\bm{f}}_{\mathrm{E}}$ on the outer side
for the particle near the wall
is smaller than that for the particle far from the wall,
which results in $\bm{F}_{\mathrm{EW}}$ directed outward.
The outward $\bm{F}_{\mathrm{EW}}$ is larger than the inward $\bm{F}_{\mathrm{E} N_1}$ near the wall,
and $\widetilde{\bm{F}}_{\mathrm{E}}\approx\bm{F}_{\mathrm{EW}}$ is directed outward there.
This behavior of $\widetilde{\bm{F}}_{\mathrm{E}}$ is enhanced
as $\mathrm{Wi}$ increases (Fig.~\ref{fig:conformationtensor}(c)).

When the particle is placed near the wall off the midline
($(y_{\mathrm{C}},z_{\mathrm{C}}) \approx (0.31,0.17)$),
strongly elongated polymers appear
in the upstream region with $y>y_{\mathrm{C}}$ and $z>z_{\mathrm{C}}$,
and they provide $\bm{F}_{\mathrm{EW}}$ directed outward
(Figs.~\ref{fig:conformationtensor}(d)--(e)).
On the other hand,
$-\nabla N_1$ and hence $\bm{F}_{\mathrm{E} N_1}$ are directed inward and toward the diagonal.
The radial components of $\bm{F}_{\mathrm{EW}}$ and $\bm{F}_{\mathrm{E} N_1}$ almost balance,
and $\widetilde{\bm{F}}_{\mathrm{E}}\approx\bm{F}_{\mathrm{EW}}+\bm{F}_{\mathrm{E} N_1}$ is directed to the diagonal.

In this Letter,
the focusing positions of spherical particles suspended in dilute PVP solutions
observed in the laboratory experiment
are reproduced by numerical computations based on the FENE-P model for the viscoelastic flow.
In the laboratory experiment with $\mathrm{Re}=50$ and $\kappa = 0.15$,
the four focusing positions appear each on the midlines and the diagonals
when $\phi=1$~wt\% and $2.5$~wt\%, respectively.
At $\phi = 1.9$~wt\% between these two concentrations,
the particles focus near these eight equilibrium positions,
indicating that
the MEP and the DEP are bistable.
The focusing positions appear as sinks of the vector field of the lift in the numerical computations,
and the transition of the focusing patterns due to the PVP concentration
was replicated by changing $\mathrm{Wi}$ in the numerical computations.
The elastic lift stems mainly from the inward lift due to the first normal stress difference
and the outward lift due to the polymer elongation near the tube wall.
The radial balance between these two major elements of the elastic lift
makes the position of the contour of $F_r=0$ independent of $\mathrm{Wi}$.
In contrast,
the closed-loop contour of $F_{\theta}=0$ shrinks,
since these two elements are stronger
and the elastic lift is directed more toward the diagonal
as $\mathrm{Wi}$ increases.
The position switch of the zero contours makes
the MEP unstable and the DEP stable.
This transition results from the interaction between the non-Newtonian properties of the PVP solution and the particle's motion.
The lateral migration of particles reported in this Letter does not originate from the secondary flow,
since the secondary flow is extremely weak due to the weak non-Newtonian properties of the medium.

\begin{acknowledgments}
We thank Dr.\ Takeshi Matsumoto for his valuable advice on numerical computations
based on the FENE-P model.
We acknowledge Katsuhiko Uno for his support on the numerical computations.
The numerical computations in this work were carried out
at Research Institute for Information Technology, Kyushu University.
This work was partially supported by KAKENHI Grant No.~20H02072 and No.~20H04504.
\end{acknowledgments}

\end{document}